\documentclass{elsart}
\usepackage{amsmath}
\usepackage{amsfonts}
\usepackage{amssymb}

\journal{Physic Letters B}
\begin{document}

\begin{frontmatter}
\title{Charge non-conservation, dequantisation, and induced electric dipole moments in varying-$\alpha$ theories}
\author{Douglas J. Shaw},
\ead{D.Shaw@damtp.cam.ac.uk}
\address{DAMTP, Centre for Mathematical Sciences, University of Cambridge, \\
Wilberforce Road, Cambridge CB3 OWA, UK}

\begin{abstract}
We note that in extensions of the Standard Model that allow for a varying fine structure constant, $\alpha$, all matter species, apart from right-handed neutrinos, will gain an intrinsic electric dipole moment (EDM).  In a large subset of varying-$\alpha$ theories, all such particle species will also gain an effective electric charge. This charge will in general not be quantised and can result in macroscopic non-conservation of electric charge.
\end{abstract}

\end{frontmatter}

\section{Introduction}
Motivated by the studies of fine-structure in the absorption lines of dust-clouds about quasars by Webb \emph{et al.}, \cite{Webb:2001}, recent years have seen a growing interest in the possibility that the fine-structure constant, $\alpha_{em} = e^2/\hbar c$, can vary in space and time.   The observations of Webb \emph{et al.} favour a value of $\alpha_{em}$ at redshifts 1-3.5 that is lower that it is today: $%
\Delta \alpha _{em}/\alpha _{em}\equiv \lbrack \alpha _{em}(z)-\alpha
_{em}(0)]/\alpha _{em}(0)=-0.57\pm 0.10\times 10^{-5}$.  A similar observational study using a different data set did not however see a variation in $\alpha_{em}$, \cite{chand1}.
There is no shortage of other astrophysical, geological, and experimental bounds on the time-variation of $\alpha_{em}$.  An excellent review of these matters has been given by Uzan in ref. \cite{uzan}.
There has also been a great deal of effort concentrated on constructing and constraining a self-consistent theoretical framework to explain $\alpha_{em}$'s apparent cosmological change  \cite{Bekenstein:1982,Sandvik:2001}.  It has been noted by several authors that, if $\alpha_{em}$ can vary, then other Standard Model gauge coupling `constants' should also be able to change, \cite{drink}.  Indeed, as a result of electroweak unification a change in $\alpha_{em}$ implies  that at least one of the weak coupling constants, $g_{W}$ and $g_{Y}$, must also change, \cite{Kimberly:2003,shaw:2004}.  
In this letter we will show that, with the exception of the cases where the ratio $g_{Y}/g_{W} := \tan \theta_{W}$ is a true constant, charge non-conservation, dequantisation and a charged neutrino are generic features of almost all varying-coupling electroweak theories. Such theories will ensure that all matter species gain an electric dipole moment (EDM).

\section{General theory}
The electroweak couplings, $g_{W}$ and $g_{Y}$, are made spacetime dependent by the definitions: $g_{W} := e^{\varphi}$ and $g_{Y} := e^{\chi}$; $\varphi(x)$ and $\chi(x)$ are scalar fields. We do not exclude the possibility that they may be functions of each other.  In general we assume $\nabla_{\mu} \varphi \neq 0$ and $\nabla_{\mu} \chi \neq 0$.  It is not necessary for what follows to say anything more specific about the dynamics of the dilaton fields, $\chi$ and $\varphi$. Gauge-invariance fixes the gauge-kinetic sector of such a theory to be:
\begin{eqnarray}
\mathcal{L}_{g} = -\tfrac{1}{4}\mathrm{tr} \, \mathcal{F}_{W \, \mu \nu} \mathcal{F}^{\mu \nu}_{W} -\tfrac{1}{4}\mathcal{F}_{Y \, \mu \nu}\mathcal{F}_{Y}^{\mu \nu}
\end{eqnarray}
\noindent where the field strengths, $\mathcal{F}_{W}^{\mu \nu}$ and $\mathcal{F}_{Y}^{\mu \nu}$ are given by:
\begin{eqnarray}
\mathcal{F}_{W}^{\mu \nu} &=& \mathbf{F}_{W}^{\mu \nu} + \partial^{\mu}\varphi \mathbf{W}^{\nu} - \partial^{\nu}\varphi \mathbf{W}^{\mu}, \\
\mathcal{F}_{Y}^{\mu \nu} &=& F_{Y}^{\mu \nu} + \partial^{\mu}\chi Y^{\nu} - \partial^{\nu}\chi Y^{\mu}. 
\end{eqnarray}
\noindent The $\mathbf{F}_{W}^{\mu \nu}$ and $F_{Y}^{\mu \nu}$ are respectively given by standard expressions for the $SU(2)$ and $U(1)$ Yang-Mills field strengths; $\mathbf{W}^{\mu}$ and $Y^{\mu}$ are the gauge fields. In all but the special case where $\varphi \equiv \chi + const$, the ratio of $g_{W}$ and $g_{Y}$, and hence $\theta_{W} := \arctan \left(g_{Y}/g_{W}\right)$ will not be constant. Moreover, as a result of renormalisation, even if $\theta_{W}$ is spacetime independent at one particular energy scale it will not be at all others.  The fine structure constant is given by:
\begin{eqnarray}
&\alpha = g^2_{W} \sin^2\theta_{W} := e^{2\phi}; \qquad \phi = \phi(\varphi, \chi).
\end{eqnarray}

\section{Charge non-conservation and simple varying-$\alpha$ theories}
It follows from Noether's principle that any gauge-invariant varying-alpha theory contains a conserved current. The class of theories described above is symmetric under modified $U(1)_{em}$ gauge transformations $A_{\mu} \rightarrow A_{\mu} + e^{-\phi}\nabla_{\mu}  \Lambda$, where $\alpha = e^{2\phi}$.  The rest of the gauge symmetry must be broken by a Higgs sector if it is to describe our universe. The conserved current however is not the one conjugate to $A_{\mu}$, i.e.  $J^{\mu}(x) := \frac{\delta \mathcal{S}_{matter}}{\delta A_{\mu}(x)}$ (with $\mathcal{S}_{matter}$ being the matter action). Noether's principle says $\nabla_{\mu} \left(e^{\phi} J^{\mu}\right) = 0$ so it is $j^{\mu} := e^{\phi}J^{\mu}$ that is conserved.  The question of which of $J^{\mu}$ and $j^{\mu}$ should be considered physical rests in the form of the dilaton-to-matter coupling.  Refs. \cite{beknew2} and \cite{shaw:2004} respectively take different stances on the issue. In `$j^{\mu}$-physical' theories charge is clearly conserved. 
 
 If $J^{\mu}$ is naturally interpreted as the physical current then there is a form of non-conservation of charge. The total charge, $Q$, in a volume $V$ is:
\begin{eqnarray}
\label{monopole}
Q :&=& \int_{V} \, j_{0} e^{\phi(t,\mathbf{x})} \, \mathrm{d}V = \int_{V} \, j_{0} \left(e^{\phi(t,\mathbf{0})} + \mathbf{x} \cdot \vec{\nabla} e^{\phi(t,\mathbf{0})} + ...\right) \\ \nonumber &=& e^{\phi(t,\mathbf{0})} q + \vec{\nabla} e^{\phi(t,\mathbf{0})} \cdot \mathbf{d} + ... + \vec{\nabla}^{i}  \vec{\nabla}^{j} ... \vec{\nabla}^{k} e^{\phi(t,\mathbf{0})} d^{(n)}_{ij..k} + ...
\end{eqnarray}
\nonumber where $d^{(n)}_{ij..k}$ is the $n^{th}$ electric multipole moment w.r.t. to the conserved current $j^{\mu}$. A collection of neutral particles cannot develop an electric charge in such theories. Similarly an initially electrically neutral, perfect fluid (containing a mixture of negatively and positively charged components) cannot become charged since all multipole moments will vanish for such a fluid. This implies that cosmologically, at least, charge will be conserved to a very good approximation.  The universe cannot develop a non-negligible overall charge in this way. Particle level interactions will also conserved charge at each vertex as a result of the conservation of $j^{\mu}$.

We will now show that when $\theta_{W}$ and $\alpha$ vary then a stronger form of non-conservation of charge arises in `$J^{\mu}$-physical' theories, and that even in `$j^{\mu}$-physical' theories  the fermions develop an EDM. 

\section{A new interaction from varying-$\theta_{W}$}
A Higgs sector must break the  $SU(2)_{L} \times U(1)_{Y}$ symmetry down to the $U(1)$ of electromagnetism. The physically propagating fields, the photon, $A^{\mu}$, and the Z-boson, $Z^{\mu}$, are given in terms of $Y^{\mu}$ and $W_{3}^{\mu}$ in the usual way. Their field strengths are:
\begin{eqnarray}
F_{A}^{\mu \nu} &=& 2e^{-1} \partial^{[\mu} \left(eA^{\nu]}\right), \\
F_{Z}^{\mu \nu} &=& 2(g_{W}\cos\theta_{W})^{-1} \partial^{[\mu} \left(g_{W}\cos\theta_{W}Z^{\nu]}\right),
\end{eqnarray}
\noindent where $e = g_{W}\sin\theta_{W} := e^{\phi}$ is the fundamental electric charge. The kinetic terms for $W^{3}$ and $Y$ now become:
\begin{eqnarray}
\mathcal{L}_{Z,A}:=&-&\tfrac{1}{4}F_{W3}^2 - \tfrac{1}{4}F_{Y}^2 = -\tfrac{1}{4}F_{A}^2 - \tfrac{1}{4}F_{Z}^2 \\ \nonumber &+& 2 F_{A}^{\mu \nu} \partial_{\mu} \theta_{W} Z_{\nu}  - 2 \tan \theta_{W} F_{Z}^{\mu \nu} \partial_{\mu} \theta_{W} Z_{\nu} \\ \nonumber &-& \tfrac{2}{\sin^2 \theta_{W} \cos^2 \theta_{W}} \left[\left(\partial \theta_{W}\right)^2 Z^2 - \partial_{\mu} \theta_{W} \partial_{\nu} \theta_{W} Z^{\mu} Z^{\nu}\right]
\end{eqnarray}
The first two terms are the standard kinetic terms for the photon and $Z$-boson. The term in square brackets and the one before provide only minor corrections to the $Z$-boson propagator. The third term, $2 F_{A}^{\mu \nu} \partial_{\mu} \theta_{W} Z_{\nu}$,  is the one that interests us.  It produces a coupling between the photon and the $Z$ boson that was not previously present.  It means that all particle species with weak neutral charge will induce an electric current density.
\section{Induced currents}
At energies well below the $Z$-boson mass, $M_{Z} \sim 91$ GeV, $Z^{\mu} \approx J^{\mu}_{N}/M_{Z}^2$; $J^{\mu}_{N}$ the weak neutral current density.  The interaction term of section 4 therefore produces an effective electromagnetic current density, $\hat{J}^{\mu}$, given by:
\begin{eqnarray}
\hat{J}^{\mu} \equiv e^{\phi} \hat{j}^{\mu} := 2e^{\phi} \nabla_{\nu} \left(\frac{\nabla^{\mu} \theta_{W} J_{N}^{\nu} - \nabla^{\nu} \theta_{W} J_{N}^{\mu} }{e^{\phi} M_{Z}^2}\right).
\end{eqnarray}

The nature of the physical electric potential depends on whether $J^{\mu}$ or $j^{\mu}$ is the physical current density. When the magnetic field vanishes, $\mathbf{B}=\mathbf{0}$, the physical potential is defined by the condition that the electric field $\mathbf{E}$ should vanish if and only if the potential is constant. When $\mathbf{B}=\mathbf{0}$, the modified Maxwell equations are:
\begin{eqnarray}
e^{\phi} \vec{\nabla} \cdot (e^{-\phi} \mathbf{E}) &=& \rho := J^{0} \\
\vec{\nabla} \times \left(e^{\phi} \mathbf{E}\right) &=& \mathbf{0}
\end{eqnarray}
So long as the gradients in $\phi$ varying only very slightly within the region of space where $\rho(x)$ has support, then $\mu = \left(\vec{\nabla} \phi\right)^2 - \nabla^2 \phi \approx const$. The electric field is given by $\mathbf{E} := e^{-\phi} \vec{\nabla}\left(e^{\phi} \Psi\right)$ where:
\begin{eqnarray}
\Psi(x) = -\frac{1}{4 \pi}\int \mathrm{d}^3 x' \mathrm{Re}\left(\frac{e^{-\sqrt{\mu}\left\vert x - x'\right \vert}}{\left \vert x - x'\right \vert}\right) \rho(x')
\end{eqnarray}
This case is the one that corresponds to $J^{\mu}$ being the physical current density. Here, $\Phi := e^{\phi}\Psi(x)$ is deemed to be the physical potential. 

This is not the only possibility.  If the $\phi$ equation of motion is such that $\vec{\nabla} \phi \times \mathbf{E} = \mathbf{0}$ whenever $\mathbf{B}=0$, then $\mathbf{E} = e^{\phi} \vec{\nabla} \Upsilon$ with:
\begin{eqnarray}
\Upsilon(x)  = -\frac{1}{4 \pi}\int \mathrm{d}^3 x'  \frac{e^{-\phi}\rho(x')}{\left \vert x - x'\right \vert}.
\end{eqnarray}
It is clear that here $e^{-\phi}\rho(x)=j^{0}$ is the physical charge density; $\Upsilon(x)$ is identified as the physical potential. 
The requirement that $\vec{\nabla} \phi \times \mathbf{E} = 0$ might seem quite contrived. It might, however, arise as an integrability condition for the $\phi$-equation of motion; for example as in ref. \cite{beknew2}.   This condition defines how the mass of any charged particle should depend on $\alpha$.  All charged particles must develop this $\alpha$-dependent mass through photon and dilaton loop corrections.  Chiral fermions are protected against becoming massive in this way, therefore all viable `$j^{\mu}$-physical' theories cannot contain charged chiral fermions. This statement applies equally to all charges associated with varying-gauge couplings.  Weakly charged neutrinos must therefore be massive in `$j^{\mu}$-physical' varying-$\alpha$ theories.

Consider a point particle, weak neutral charge $Q_{N}$, at $\mathbf{x}=0$.  In a `$J^{\mu}$-physical' theory the new interaction term described above makes the following contribution to the physical electric potential $\Phi(x)$:
\begin{eqnarray}
\Phi(x) \approx \left(-\frac{Q_{N}\vec{\nabla} \theta_{W} \cdot \vec{\nabla} \phi}{M_{Z}^2}\right) \frac{e^{\phi(x)}}{2\pi r} + \frac{Q_{N}\vec{\nabla} \theta_{W}}{M_{Z}^2} \cdot \frac{\mathbf{x}e^{\phi(x)}}{2 \pi r^3},
\end{eqnarray}
\noindent where $r = \left\vert \mathbf{x} \right\vert$. The first term in $\Phi$ represents a point electric charge $q_{eff} = \frac{2Q_{N} \vec{\nabla} \theta_{W} \cdot \vec{\nabla} \phi }{M_{Z}^2}$.
The second term is the potential of an electric-dipole moment $\mathbf{d}_{eff} = -\frac{2Q_{N}\vec{\nabla} \theta_{W}}{M_{Z}^2}$.  In `$J^{\mu}$-physical' theories all weak neutrally-charged particles will become effectively electrically charged when $\theta_{W}$ varies.  Such particles will also develop an effective EDM. The form of $q_{eff}$ means that it will not be quantised in units of $e$. There is  effective dequantisation of electric charge in these theories. In `$j^{\mu}$-physical' theories we do not see an induced charge effect. Weak neutrally-charged particles will still develop an EDM. The electric potential, $\Upsilon(x)$, is:
\begin{eqnarray}
\Upsilon(x) = \frac{Q_{N}\vec{\nabla} \theta_{W}}{e^{\phi(0)}M_{Z}^2} \cdot \frac{\mathbf{x}}{2 \pi r^3}.
\end{eqnarray}
The induced EDM is
\begin{eqnarray}
\mathbf{d}'_{eff} := -\frac{2Q_{N}\vec{\nabla} \theta_{W}}{e^{\phi(0)M_{Z}^2}}.
\end{eqnarray}
Order of magnitude estimates for $q_{eff}$, $\mathbf{d}_{eff}$ and $\mathbf{d}'_{eff}$ are given in the section 6 below.
\section{Discussion}
The weak neutral current, $J_{N}^{\mu}$, is not conserved. In general, massive bodies such as our Sun, and the universe as a whole, have a large net weak-neutral charge density compared to their net electric charge density.   In `$J^{\mu}$-physical' electroweak theories of varying-$\alpha$, particles develop an electric charge proportional to their weak-neutral charge.  It is possible then to have non-conservation of electric charge. The universe develops a non-negligible overall charge in this way.  The charges that are induced are in general not quantised in units of the fundamental charge $e$.   Spatial variations in $\theta_{W}$ also induce EDMs on the fundamental fermion species. The EDMs all point in the direction of  $\vec{\nabla} \theta_{W}$.  In a region where $\vec{\nabla} \theta_{W} \approx const$, therefore, these EDMs will line up and produce an overall macroscopic EDM. Numerically the sizes of the effective charges and EDMs are:
\begin{eqnarray}
q_{eff} &\sim& 10^{-31} e  \left(\left \Vert \nabla \theta_{W} \right \Vert \cdot \left \Vert \nabla \ln \alpha \right \Vert \mathrm{cm}^{2}\right) \\
d_{eff} &\approx& d'_{eff} \sim 10^{-31} e\mathrm{-cm} \left(\left \Vert \nabla \theta_{W} \right \Vert \mathrm{cm}\right)
\end{eqnarray}
In many varying-$\alpha$ theories one finds $\vec{\nabla} \ln \alpha \approx \zeta_{\alpha} \vec{\nabla} \phi_{N}$, $\vec{\nabla} \theta_{W} \approx \zeta_{\theta} \vec{\nabla} \phi_{N}$, where $\phi_{N}=GM/r^2$ is the Newtonian gravitational potential. We expect $\zeta_{\alpha},\zeta_{\theta} \ll 1$. Near the surface of Earth such theories would induce $q_{eff}  \sim \zeta_{\alpha} \zeta_{\theta} \, 10^{-66} e$, $d_{eff} \approx d'_{eff} \sim \zeta_{\theta} \, 10^{-48}$ \emph{e}-cm. 

Any physically viable, varying-$\alpha$ and varying-$\theta_{W}$ theory must satisfy all relevant bounds on the neutrino and neutron charges and on the EDMs of the fundamental particles. The most restrictive upper bound on the electron-neutrino charge, $q_{\nu}$, has been given by  Caprini and Ferreira in ref. \cite{Caprini}. They considered the isotropy of the Cosmic Microwave Background (CMB) and found:  $q_\nu<4\times10^{-35}e$.  In the same way they also bounded the charge difference between a proton and an electron: $q_{e-p} < 10^{-26} e$. An upper bound on the neutron charge, $q_{n} < -0.4\pm1.1 \times 10^{-21}$, is given by Baumann \emph{et al.} in \cite{Baumann}.  The Particle Data Group, see ref. \cite{pdg}, gives the upper bound on the electron EDM as $d_{e} < 6.9 \pm 7.4 \times 10^{-28}$ \emph{e}-cm.  Experiments are planned that would be able to detect any electron EDM at the $10^{-31}$ \emph{e}-cm level, \cite{Kawall}.   Ref. \cite{pdg} also gives upper bounds on the EDMs of the proton, $d_{p} < 0.54 \times 10^{-23}$ \emph{e}-cm, and the neutron, $d_{n} <  0.63 \times 10^{-25}$ \emph{e}-cm.  It is clear that all current bounds will be easily satisfied by most varying-$\alpha$ theories.

It is normally the case that intrinsic EDMs on Dirac fermions are indicators of CP-violation.  In varying-$\alpha$ theories we have seen that it possible to induce such EDMs without adding any explicit CP-violating term to the Lagrangian and that varying-$\alpha$ theories generically result in some manner of charge non-conservation and effective dequantisation of charge without breaking the $U(1)_{em}$ symmetry.  These effects, if detectable in the context of a given theory, could provide us with a new way of probing the rate of spatial variation in $\theta_{W}$ and $\alpha$.

\textbf{Acknowledgements} DS is supported by a PPARC studentship. I would
like to thank J. D. Barrow for reading a preprint of this letter and for making helpful suggestions, and to thank J. Magueijo for helpful discussions.

\end{document}